\documentclass{PoS}
\usepackage{graphicx}

\PoS{PoS(LAT2005)111}
\title{The PACS-CS Project}
\ShortTitle{The PACS-CS Project }
\author{PACS-CS Collaboration : S. Aoki${}^{a,b}$, K.-I. Ishikawa${}^c$, T. Ishikawa${}^d$, N. Ishizuka${}^{a,d}$, K. Kanaya${}^a$, Y. Kuramashi${}^{a,d}$, M. Okawa${}^c$, K. Sasaki${}^d$, Y. Taniguchi${}^{a,d}$, N. Tsutsui${}^e$, \speaker{A. Ukawa${}^{a,d}$}\thanks{ukawa@ccs.tsukuba.ac.jp}, T. Yoshi\'e${}^{a,d}$\\
${}^a$Graduate School of Pure and Applied Sciences, University of Tsukuba, Tsukuba, Ibaraki 305-8571, Japan\\
${}^b$Riken BNL Research Center, Brookhaven National Laboratory, Upton, New York 11973, USA\\
${}^c$Department of Physics, Hiroshima University, Higashi-Hiroshima, Hiroshima 739-8526, Japan\\
${}^d$Center for Computational Sciences, University of Tsukuba, Tsukuba, Ibaraki 305-8577, Japan\\
${}^e$High Energy Accelerator Research Organization (KEK), 
Tsukuba, Ibaraki 305-0801, Japan}

\abstract{
We describe our plan to develop a large-scale cluster system 
with a peak speed of
14.3Tflops for lattice QCD at the Center for Computational Sciences,
University of Tsukuba, as a successor to the current 0.6Tflops CP-PACS
computer. 
The system consist of 2560 nodes connected by a 16x16x10 three-dimensional 
hyper crossbar network.  Each node has a single low-voltage 2.8GHz Xeon 
processor and 2GBytes of memory with 6.4GBytes/sec bandwidth, 
and 160 GBytes of disk in RAID1 mode.  The network link in each of the 
three directions is made of dual Gigabit Ethernet with the peak throughput 
of 250MByte/sec.  Hence each node has an aggregate network bandwidth of 
750MByte/sec.  The system will run under Linux and SCore, 
and an extension of the PM driver is developed for the network.  
The system will be developed jointly with Hitachi Limited.  
The installation is scheduled in the first quarter of Japanese Fiscal 2006
(April-June 2006) and the start of operation is expected in July 2006.
}

\FullConference{XXIIIrd International Symposium on Lattice Field Theory\\
		 25-30 July 2005\\
		 Trinity College, Dublin, Ireland}

\begin{document}

\section{Introduction}

Progress in lattice QCD requires a dual research program of 
pursuing physics with computers available at the time, and 
developing computers themselves aiming at the next stage. 
In this article, we describe our project to develop a successor to the 
current CP-PACS system\cite{cppacs} for lattice QCD at the Center for 
Computational Sciences, University of Tsukuba.

The University of Tsukuba has a long tradition of developing parallel 
computers for scientific applications, dating back to PACS-9
built by T. Hoshino and his collaborators in 1978.  
Over the years, successively more powerful systems have been developed as 
shown in Table~\ref{tab:pacs-machines}.   Lattice QCD became the main 
application with QCDPAX, and the trend continued with the 6th generation 
CP-PACS built almost 10 years ago.  

All the systems built so far carries either PACS (Processor Array for 
Continuum Simulation or Parallel Array Computer System) or PAX 
(Parallel Array Experiment or Processor Array Experiment) in their naming.  
The next system, currently under development and will be the 7th of the 
series, is named 
{\it PACS-CS - Parallel Array Computer System for Computational 
Sciences - } as the target applications encompass density-functional 
simulations in addition to fundamental physics ones including lattice QCD. 

In this article, we discuss the design considerations, the machine 
specification including hardware and software, the performance 
benchmarks available at present, and the development schedule, 
of the PACS-CS computer\cite{pacs-cs}.

\begin{table}[b]
\begin{center}
\begin{tabular}{lll}
\hline
year&name&peek speed\\
\hline
1978  &PACS-9  &7 kflops\\
1980  &PAXS-32 &500 kflops\\
1983  &PAX-128 &4 Mflops\\
1984  &PAX-32J &3 Mflops\\
1989  &QCDPAX  &14 Gflops\\
1996  &CP-PACS &614 Gflops\\
2006  &PACS-CS &14.3 Tflops\\
\hline
\end{tabular}
\end{center}
\label{tab:pacs-machines}
\caption{PAX/PACS series of parallel computers developed at University of 
Tsukuba}
\end{table}

\begin{figure}[t]
\begin{center}
\includegraphics[width=100mm,angle=0]{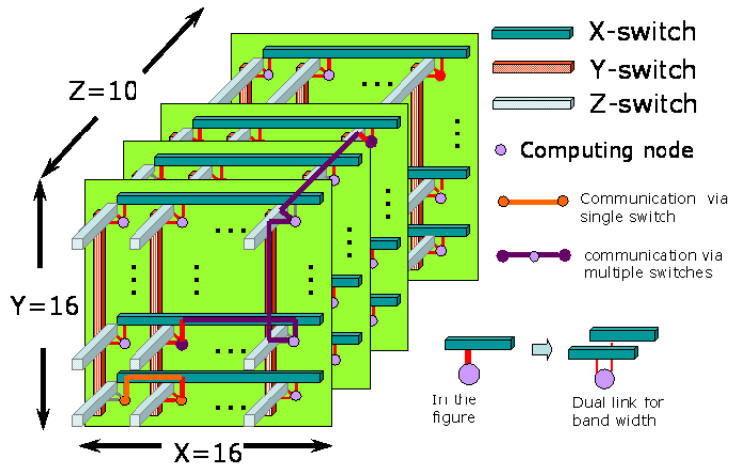}
\end{center}
\caption{Schematic view of the 3-d hyper-crossbar network of the 
PACS-CS system.}
\label{fig:pacs-cs}
\end{figure}

\begin{figure}[t]
\begin{center}
\includegraphics[width=100mm,angle=0]{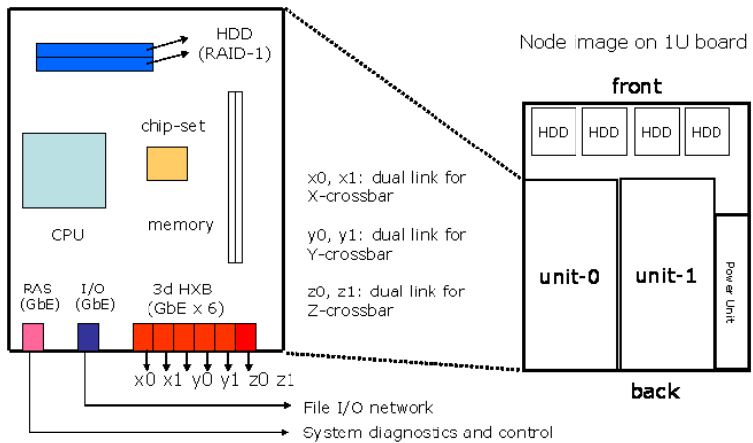}
\end{center}
\caption{Schematic view of mother board design.}
\label{fig:motherboard}
\end{figure}

\section{Design considerations and system specification}

For a variety of reasons including keeping the development period as short as 
possible as well as cost, we have decided to pursue a 
commodity approach to build the next computer system for our lattice QCD 
program.  Standard clusters which have become widely available over 
the last several 
years, however, are unsatisfactory in the following points: 
(i) with the 
standard SMP configuration in which each node is equipped with dual 
or more processors, the memory to processor bandwidth is very low so that 
only a fraction of the peak performance is actually realized, and similarly   
(ii) the tree-type interconnect with a small number of large switches is 
problematical in that the bandwidth is too low if inexpensive Gigabit Ethernet 
is used, while the cost, particularly that of switches,  increases rapidly 
if a faster interconnect such as Myrinet or Infiniband is adopted. 
Our choice to resolve these issues has led to the current PACS-CS design, 
which is essentially an MPP system built out of commodity components. 

\begin{table}[b]
\begin{center}
\begin{tabular}{ll|lll}
\hline
Number of nodes & 2560& \multicolumn{3}{l}{Node configuration}\\
Peak performance & 14.3 Tflops &CPU  &single LV Xeon 2.8GHz  &5.6Gflops\\
Total memory & 5 TByte    &Memory  &\multicolumn{2}{l}{2 GByte at 6.4GByte/s bandwidth}\\
Total disk space & 0.41 PByte &local disk  &\multicolumn{2}{l}{160 GByte$\times 2$ (RIAD-1 mirror)}\\
Interconnect& $16\times 16\times 10$&\multicolumn{3}{l}{Interconnect bandwidth} \\
&3-dim hyper-crossbar &&250MByte/s$\times 2$/link\\
OS  & Linux and SCore& &750MByte/s$\times 2$/node\\
Programming & Fortran90, C, C++, MPI\\
\hline
System size & 59 racks & Estimated power & 545 kW\\
\hline 
\end{tabular}
\end{center}
\label{tab:pacs-cs}
\caption{PACS-CS design specification}
\end{table}
   
We equip each node with a single processor and connect it  
with the fastest bus available to the main memory with a matched I/O 
speed.  Specifically, we choose FSB800 for the bus and two banks of PC3200 
DDR2 memory, which means the memory to processor bandwidth of 6.4GByte/s.  
For the processor, we use Intel Xeon for a better error check and correction 
than the Pentium series, 
with the frequency of 2.8GHz (5.6Gflops peak) since higher rates will not 
provide higher effective performance, and the low-voltage version to 
suppress power consumption. 

For the interconnect, we adopt the three-dimensional hyper-crossbar network 
used for the CP-PACS computer.  This network provides a versatile 
3-dimensional connection, much more flexible than the mesh, with a medium-sized
switches, albeit large in number, and so helps reduce the switch cost.
  
In Fig.~\ref{fig:pacs-cs} we show a schematic view of the hyper-crossbar 
network of the PACS-CS system.  The actual system has 
2560 nodes arranged in a $16\times 16\times 10$ topology. 
The network link in each direction consists of dual Gigabit Ethernet. 
With the trunking of two links of Ethernets, the peak bandwidth 
equals 250MByte/s for each of the three directions, 
and the total bandwidth of each node sums up to 
750MByte/s for each node.  This is competitive with 
high-throughput interconnects such as InfiniBand which allows 1 GByte/s  
bandwidth. 

In order to keep the packaging density equal to the standard dual SMP cluster, 
we put two nodes on a single 1U board.  Each node has to have at least 
6 Gigabit Ethernet ports.  In fact we put two more ports, one for I/O and the 
other for system diagnostics and control.  In addition, the chipset is 
carefully arranged so that a sufficient bandwidth is ensured 
between memory and each of the six Gigabit Ethernet Interface. 
A schematic view of our mother borad is given in Fig.~\ref{fig:motherboard}.

For temporary data storage, 
each node is equipped with a 160 GByte of disk space, which
is duplicated to work in the RAID-1 mode.  
The external I/O is made via a three-stage Gigabit Ethernet tree whose 
strength is doubled at each upward step.  The file server for I/O is 
connected to a 10 TByte RAID-5 disk. 

The operating system of PACS-CS is LINUX, and SCore~\cite{score} is 
used as the cluster middleware.  
We need to develop the  driver for data communication between 
nodes over the hyper-crossbar network.  Work is in progress\cite{pm-hxb} 
based on the PMv2 driver available from SCore.

The programming language is Fortran90, C and C++.  Communication 
is handled by MPI which will call the hyper-crossbar driver 
explained above. 

In Table~\ref{tab:pacs-cs} we list the current design specification of the 
PACS-CS computer.  
 
\section{Benchmark}

The primary issue with the system performance is the effective floating 
point performance of each node.  
We have built a test system to the specification of 
PACS-CS, {\it i.e,} with the same processor, chipset and 
memory components.  We have tested the MULT benchmark code written and 
optimized by K. Ishikawa\cite{mult} which measures the node performance for 
the multiplication by the Wilson-clover hopping matrix given by 
\begin{equation}
\left( 1 + c_{sw}F\cdot\sigma\right)^{-1} 
\sum_\mu\left((1-\gamma_\mu) U_{n\mu}+(1+\gamma_\mu)U^\dagger_{n\mu}\right)
\end{equation}
Using the Intel C and Fortran compiler version 8.1 allowing the use of EM64T 
features, this code achieved over 30\% of the peak speed as summarized in 
Table \ref{tab:flops}.  

The second issue with the system performance is the network.
The network driver being developed is called PM/Ethernet-HXB\cite{pm-hxb}. 
The driver is designed to enable trunking of up to 8 Ethernet links, routing 
of messages over a multi-dimensional crossbar, and is implemented with the 
zero-copy communication feature avoiding system buffering.  
The performance figures are available at present only at the driver 
level, which is listed in Table \ref{tab:hxb} where a dual trunking 
is assumed for each direction to follow the PACS-CS design.    
The throughput reaches almost the peak capacity.  The latency, as expected, 
is sizable, and increases proportionately with the number of dimension.  
Further tests at the MPI level, and with actual lattice QCD codes, 
will be made soon.  

\begin{table}[b]
\begin{center}
\begin{tabular}{lll}
\hline
coding & Gflops & \% of peak (5.6Gflops)\\
\hline
C with SSE3 assembler coding & 1.87 & 33\%\\
C with Intel intrinsic function & 1.91 & 34\%\\
Fortran & 1.45 & 26\% \\
\hline 
\end{tabular}
\end{center}
\caption{Performance for MULT benchmark for the lattice size $8\times 8\times 8\times 64$}
\label{tab:flops}
\end{table}
   
\begin{table}
\begin{center}
\begin{tabular}{lll}
\hline
  & max bandwidth & latency\\
\hline
1 dimension & 247 MByte/s (99\%) & 15.1 $\mu$s\\
2 dimension & 241 MByte/s (96\%) & 29.1 $\mu$s\\
3 dimension & 237 MByte/s (95\%) & 43.2 $\mu$s\\
\hline 
\end{tabular}
\end{center}
\caption{Driver-level performance of the PM/Ethernet-HXB for 3-d 
hyper-crossbar network for dual trunking in each dimension.
\protect\cite{pm-hxb}}
\label{tab:hxb}
\end{table}
   
\section{System implementation}

The system is housed in the standard 42U 19-inch racks.  Two processor nodes 
are placed on an 1U board, 32 boards are placed in a rack.  
There will be 40 racks for the planned 2560 node system.  
A 48-port Gigabit Ethernet switch is placed on a board, and 19 racks 
will be needed to house all the switches.   
In total the full system will consist of 59 racks spread over an 
area of about 100$m^2$.
The estimated power consumption is 545 kW when the system is in full 
operation. 

\section{Summary}

In this article, we have described the present status of the PACS-CS 
Project.  The Project has been approved by the Japanese Government and 
has officially started in April 2005.  Through an official bidding
procedure, Hitachi Ltd. has been selected for the development of 
the system in July 2005.  Separately, Fujitsu Ltd. has been 
chosen in August 2005 for developing the hyper-crossbar network driver.  

Currently we envisage the installation of the system in the first quarter of 
the Japanese fiscal 2006 which starts in April, and the start of operation 
for physics runs in July.  The first physics project we wish to pursue is 
the $N_f=2+1$ full QCD program using the Wilson-clover quark action 
and the Iwasaki RG-improved gluon action\cite{tomomi}, hopefully lowering the 
light quark masses below $m_\pi/m_\rho\approx 0.4$ so that chiral 
extrapolation can be brought under control.  We plan to employ the 
recent domain-decomposition acceleration ideas\cite{luescher} 
to achieve this goal.  

\section*{Acknowledgements}
This work is supported in part by the Grants-in-Aid of the Ministry of 
Education 
(Nos. 13135204, 13640260, 14046202, 15204015, 15540251,
      15740134, 16028201, 16540228, 17340066, 17540259).

\end{document}